\documentclass[preprint,12pt]{elsarticle}

\usepackage{amssymb}

\journal{Physica A}

\begin{document}

\begin{frontmatter}



\title{An introduction to Monte Carlo methods}


\author{J.-C. Walter}
\address{Laboratoire Charles Coulomb UMR 5221 $\&$ CNRS, Universit\'e Montpellier 2,\\ 34095 Montpellier, France}
\ead{jean-charles.walter@univ-montp2.fr}
\author{G. T. Barkema}
\address{Institute for Theoretical Physics, Utrecht University, The Netherlands}
\address{Instituut-Lorentz, Universiteit Leiden, P.O. Box 9506, 2300 RA Leiden, The Netherlands}
\ead{g.t.barkema@uu.nl}
\begin{abstract}
Monte Carlo simulations are methods for simulating statistical systems. 
The aim is to generate a representative ensemble of configurations to
access thermodynamical quantities without the need to solve the system
analytically or to perform an exact enumeration.
The main principles of Monte Carlo simulations are ergodicity and detailed balance.
The Ising model is a lattice spin system with nearest neighbor interactions
that is appropriate to illustrate different examples of Monte Carlo simulations.
It displays a second order phase transition between a disordered (high temperature)
and ordered (low temperature) phases, leading to different strategies of simulations.
The Metropolis algorithm and the Glauber dynamics are efficient at high temperature.
Close to the critical temperature, where the spins display long range correlations,
cluster algorithms are more efficient. We introduce the rejection free (or continuous time)
algorithm and describe in details an interesting alternative representation of the Ising
model using graphs instead of spins with the Worm algorithm. We conclude with an important 
discussion of the dynamical effects such as thermalization and correlation time.
\end{abstract}


\end{frontmatter}


\section{Introduction}
\label{intro}
Most models in statistical physics are not solvable analytically,
and therefore an alternative way is needed to determine thermodynamical
quantities. Numerical simulations help in this task, but introduce another
challenge: it is not possible, in most cases, to enumerate all the
possible configurations of a system; one therefore has to create a set of
configurations that are representative for the entire ensemble. In this section,
we will illustrate
our purpose with the Ising model. This is a renowned model because of
its simplicity and success in the description of critical phenomena~\cite{Binney95}. The
degrees of freedom are spins $S_i=\pm1$ placed at the vertex $i$ of
a lattice. This lattice will be square or cubic for simplicity, with edge size
$L$ and dimension $D$. Thus, the system contains $N=L^D$
spins. The hamiltonian of the Ising model is:
\begin{equation}
 \mathcal H=-J\sum_{\langle ij\rangle}S_iS_j\,,\label{HIsing}
\end{equation}
where the summation runs over all pairs of nearest-neighbor spins $\langle ij\rangle$ of
the lattice and $J$ is the strength of the interaction. The statistical
properties of the system are obtained from the {\it partition function}:
\begin{equation}
 Z=\sum_{\mathcal C} e^{-\beta  E({\mathcal C})}\,,
\end{equation}
where the summation runs over all the configurations $\mathcal C$. The energy of a configuration is 
denoted by $E({\mathcal C})$. Here, $\beta\equiv 1/(k_B T)$ is the inverse temperature
(temperature $T$ and Boltzmann constant $k_B$). The Ising model displays a
second-order phase transition at the temperature $T_c$, characterised by 
a high temperature phase with an average magnetization zero (disordered phase)
and a low temperature phase with a non-zero average magnetization (ordered phase). The
system is exactly solvable in one and two dimensions. For $D\ge4$, the critical properties are easily 
obtained by the renormalization group. In three dimensions no
exact solution is available. Even a $3D$ cubic lattice of very modest
size $10\times 10\times 10$ generates $2^{1000}\approx10^{301}$
configurations in the partition function. If we want to obtain e.g.
critical exponents with a sufficient accuracy, we need sizes that are at least an
order of magnitude larger. 
An exact enumeration is a hopeless effort. {\it Monte Carlo
simulations} are one of the possible ways to perform a sampling of
configurations. This sampling is made out of a set of configurations of
the phase space that contribute the most to the averages, without the
need of generating every single configuration. This is referred to as {\it
importance sampling}. In this sampling of the phase space, it is 
important to choose the appropriate Monte Carlo scheme to reduce the
computational time. In that respect, the Ising model is interesting 
because the different regimes in temperature lead to the development of new algorithm
that reduce tremendously the computational time, specifically close to the critical
temperature.

We will start these notes by introducing two important principles of
Monte Carlo simulations: detailed
balance and ergodicity. Then we will review different examples of Monte Carlo
methods applied to the Ising model: local and cluster algorithms,
the rejection free (or continuous time) algorithm, and another kind of
Monte Carlo simulations based on an alternative representation of the spin system, 
namely the so-called worm algorithm. We continue with discussing dynamical quantities,
such as the thermalization and correlation times.
\section{Principles of MC simulations: Ergodicity $\&$ detailed balance condition}
The basic idea of most Monte Carlo simulations is to iteratively propose
a small random change in a configuration $C_i$, resulting in the trial
configuration $C_{i+1}^t$ (the index ``$t$'' stands for trial). Next,
the trial configuration is either accepted, i.e. $C_{i+1}=C_{i+1}^t$,
or rejected, i.e. $C_{i+1}=C_i$. The resulting set of configurations
for $i=1 \dots M$ is known as a Markov chain in the phase space of the
system. We define $P_A(t)$ as the probability to find the system in
the configuration $A$ at the time $t$ and $W(A\to B)$ the transition
rate from the state $A$ to the state $B$. This Markov process can be
described by the master equation:
\begin{equation}
\frac{dP_A(t)}{dt}=\sum_{A\ne B}\left[P_B(t)W(B\to A)-P_A(t)W(A\to B)\right]\,,\label{meq} 
\end{equation}
with the condition $W(A\to B)\ge0$ and $\sum_B W(A\to B)=1$ for all $A$
and $B$. The transition probability $W(A\to B)$ can be further decomposed
into a trial proposition probability $T(A\to B)$ and an acceptance
probability $A(A\to B)$ so that $W(A\to B)=T(A\to B)\cdot A(A\to B)$.
A proposed change in the configuration is usually referred to as a Monte
Carlo move. Conventionally, the time scale in Monte Carlo simulations
is chosen such that each degree of freedom of the system is proposed to change
once per unit time, statistically.

The first constraint on this Markov chain is called {\it ergodicity}:
starting from any configuration $C_0$ with nonzero Boltzmann weight, any
other configuration with nonzero Boltzmann weight should be reachable
through a finite number of Monte Carlo moves. This constraint is necessary
for a proper sampling of phase space, as otherwise the Markov chain will
be unable to access a part of phase space with a nonzero contribution
to the partition sum.

Apart from a very small number of peculiar algorithms, a second constraint is known as the
{\it condition of detailed balance}. For every pair of states $A$ and $B$, the
probability to move from $A$ to $B$, as well as the probability for the reverse
move, are related via: 
\begin{equation}
P_A \cdot T(A\rightarrow B) \cdot A(A\rightarrow B) = P_B \cdot T(B\rightarrow A) \cdot A(B\rightarrow A)\,.\label{detbal}
\end{equation}
The meaning of this condition can be seen in Eq.~(\ref{meq}): a stationary
probability (i.e. $dP_A/dt=0$) is reached if each individual term in the summation on the right
hand side cancels. This prevents the Markov chain to be trapped
in a limit cycle~\cite{Newman99}. This is a strong, but not necessary,
condition. We mention that generalizations of Monte
Carlo process that do not satisfy detailed balance exist.
The combination of ergodicity and detailed balance assures a correct
algorithm, i.e., given a long enough time, the desired distribution
probability is sampled.

The key question in Monte Carlo algorithms is which small changes one
should propose, and what acceptance probabilities one should choose.
The trial proposition and acceptance probabilities have to be well
chosen so that the probability of sampling of a configuration $A$
(after thermalization) is equal to the {\it Boltzmann weight}:
\begin{equation}
 P_A = \frac{e^{-\beta E_A}}{Z}\,,\label{bolweight}
\end{equation}
in which $E_A$ is the energy of configuration $A$. The knowledge of
the partition function $Z$ is not necessary because the transition
probabilities are constructed with the ratio of probabilities. The
detailed balance condition~(\ref{detbal}), using (\ref{bolweight}),
can be rewritten as:
\begin{equation}
\frac{T(B\rightarrow A) \cdot A(B\rightarrow A)}{T(A\rightarrow B) \cdot A(A\rightarrow B)}=\frac{P_A}{P_B}=e^{-\beta(E_A-E_B)}\,.\label{detbalbol1}
\end{equation}
\section{Local MC algorithms: Metropolis \& Glauber}
One often-used approach to realize detailed balance is to propose randomly
a small change in state $A$, resulting in another state $B$, in such a
way that the reverse process (starting in $B$ and then proposing a small
change that results in $A$) is equally likely. More formally, a process
in which the condition $T(A\rightarrow B) = T(B\rightarrow A)$ holds for
all pairs of states $\{ A,B\}$. For example, taking the example of an
Ising model containing $N$ spins, it corresponds to chose randomly one
of the spins on the lattice, therefore $T(A\rightarrow B) = T(B\rightarrow
A)=1/N$. Detailed balance allows for a common scale factor in the acceptance
probabilities for the forward and reverse Monte Carlo moves, but being probabilities,
they cannot exceed 1. Simulations are then fastest if the bigger of the two acceptance probabilities
is equal to 1, i.e. either $A(A\rightarrow B)$ or  $A(B\rightarrow A)$ is equal to 1. These
conditions (including detailed balance) are realized by the so-called {\it
Metropolis algorithm}, in which the acceptance probability is given by:
\begin{equation}
A_{\rm met}(A\rightarrow B) ={\rm Min} \left[1,{P_B}/{P_A} \right]= {\rm Min} \left[1,\exp(-\beta(E_B-E_A) \right].\label{metro}
\end{equation}
Thus, a proposed move that does not raise the total energy is always
accepted, but a proposed move which results in higher energy is
accepted with a probability that decreases exponentially with the increase of the
energy difference. For the sake of illustration, let us describe how a simulation of the Ising model
looks like:
\begin{enumerate}
 \item Initialize all spins (either random or all up)
 \item Perform $N$ random trial moves ($N=L^D$):
  \begin{enumerate}
   \item randomly select a site
  \item compute the energy difference $\Delta E=E_B-E_A$ if the trial (here a spin flip) induces a change in energy
  \item generate a random number $\rm{\bf{rn}}$ uniformly distributed in $[0,1]$
  \item if $\Delta E<0$ or if $\rm{\bf{rn}}< \exp(-\Delta E)$: flip the spin
  \end{enumerate}
  \item Perform sampling of some observables
\end{enumerate}
The step 2 corresponds to one unit time step of the Monte Carlo
simulation. An alternative to the Metropolis algorithm is the
{\it Glauber dynamics}~\cite{Glauber63}. The trial probability is the same
as Metropolis i.e. $T(A\rightarrow B) = T(B\rightarrow A)=1/N$. However
the acceptance probability is now:
\begin{equation}
 A_{\rm gla}(A\to B)=\frac{e^{-\beta(E_B-E_A)}}{1+e^{-\beta(E_B-E_A)}}\,,\label{glauber}
\end{equation}
which also satisfies the detailed balance condition Eq.(\ref{detbalbol1}).
\section{Cluster algorithms: the example of the Wolff algorithm}
\label{wolff}

\begin{figure}[!ht]
\begin{center}
\centerline{\hbox{
\includegraphics[width=0.9\columnwidth]{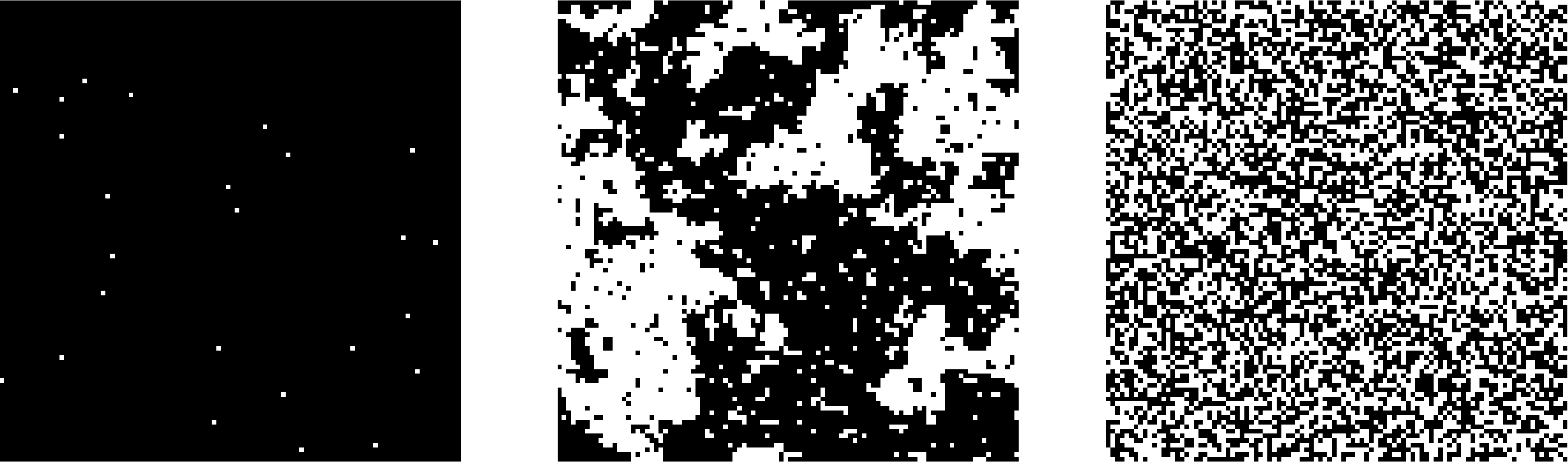}
}}
\caption{\label{Ising} Snapshots of the 2D Ising
model defined in Eq.~(\ref{HIsing}) at three different
temperatures: from left to right, $T\ll T_c$, $T\approx
T_c$ and $T\gg T_c$ where $T_c$ is the critical
temperature. White and blacks dots denote spins up and
down. The system size is $200\times200$. In the picture at
$T_c$ (middle), we observe large clusters of correlated
spins: these are {\it critical fluctuations} that slow
down Monte Carlo simulations when local algorithms such as
Metropolis or Glauber are used. This critical slowing
down is reduced by non-local (or cluster) algorithms
like the Wolff algorithm~\cite{Wolff89}.}
\end{center}
\end{figure}

Many models encounter phase transitions at some critical temperature.
The paradigmatic example for the second order phase transitions is the Ising
model defined in Eq.~(\ref{HIsing}). In the vicinity of the critical 
temperature, the spins display {\it critical fluctuations}. As shown
in Figure~\ref{Ising} (middle), large aligned spin domains appear. This phenomenon is
associated with (i) the divergence of the correlation length $\xi$
of the connected spin-spin correlation function $C(|i-j|)=\langle
S_i\cdot S_j\rangle-\langle S_i\rangle^2$ (ii) the divergence of the
 correlation time of the autocorrelation
function $C(|t-t'|)=\langle S_i(t)\cdot S_i(t')\rangle-\langle
S_i(t)\rangle^2$. Moreover, the correlation time increases with the
size of the system like $\tau\sim L^{z_c}$ where $z_c$ is
the {\it critical dynamical exponent}. For the 2D Ising model simulated
with the Metropolis algorithm, $z_c=2.1665(12)$~\cite{Nightingale96}. This
phenomenon of critical slowing down reflects the difficulty to change
the magnetization of a correlated spin cluster. Take again the example
of a 2D spin system where one spin has four nearest neighbors. If
this spin is surrounded by aligned spins, its contribution to the energy is $E_A=-4J$ and
after the reversal of this spin, this becomes $E_B=4J$. Right at
$T_c\approx2.269$, the acceptance probability is low for the Metropolis
algorithm: $A(A\to B)=e^{-8\beta_cJ}=0.0294...$. Thus, most of the flipping
attempts are rejected. Making matters worse, even if such a spin with aligned 
neighbours is flipped, the next time it is selected, it will surely flip
back. Only spin flips at the edge of a cluster
have a significant effect over a longer time; but their fraction becomes
vanishingly small when the critical temperature is approached and the cluster size
diverges.

\begin{figure}[!ht]
\begin{center}
\centerline{\vbox{
\includegraphics[width=0.8\columnwidth]{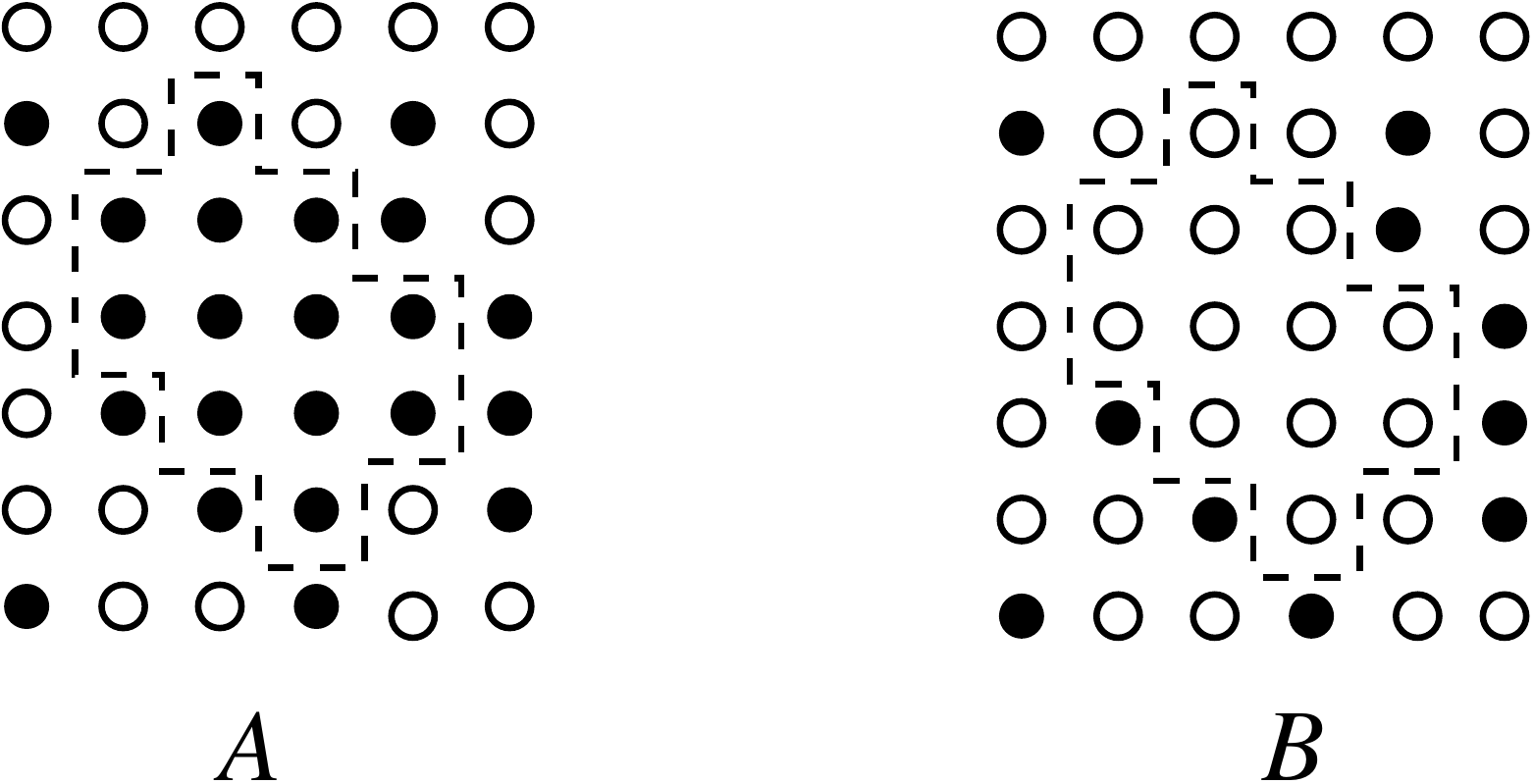}
}}
\caption{\label{Wolff} Sketch of one iteration of the non-local algorithm
introduced by Wolff~\cite{Wolff89} between two spin configurations $A$ and
$B$. The white and black dots stand for spins of opposite signs. The
spins within the loop (dashed line) belong to the same cluster. The
steps to form the cluster are: (i) choose randomly a seed spin (ii)
add aligned spins with the probability $P_{\rm add}$ (see text) (iii) add
iteratively aligned neighbors of newly added spins with the probability
$P_{\rm add}$ (iv)  flip all the spins in the cluster at once when the
cluster is completed. This is an efficient algorithm for the Ising model
at criticality.}
\end{center}
\end{figure}

One remedy is to develop a non-local algorithm
that flips a whole cluster of spins at once. Such an algorithm has been
designed for the Ising model by Wolff~\cite{Wolff89}, following the idea of
Swendsen and Wang~\cite{Swendsen87} for more general spin systems. A sketch of this procedure is shown
in Figure~\ref{Wolff}.

The procedure consists of first choosing a random initial site (seed site).
Then, we add each neighboring spin, provided it is aligned, with
the probability $P_{\rm add}$. If it is not aligned, it cannot
belong to the cluster. This step is iteratively repeated with each
neighbor added to the cluster. When no neighbor can be added to the cluster anymore, all the
spins in the cluster flip at once. The probability to form a certain cluster of
spins in state $A$ before the Wolff move is the same as that in state $B$ after the Wolff move,
except for the aligned spins that have not been added to the cluster at
the boundaries. The probability to not add an aligned spin is $1-P_{\rm
add}$. If $m$ and $n$ stand for non-added aligned spins to the cluster for
$A$ and $B$, $T(A\rightarrow B)/T(B\rightarrow A)=(1-P_{\rm add})^{m-n}$
and the detailed balance condition (\ref{detbalbol1}) can be rewritten as:
\begin{equation}
\frac{T(A\rightarrow B) \cdot A(A\rightarrow B)}{T(B\rightarrow A) \cdot A(B\rightarrow A)}
=(1-P_{\rm add})^{m-n}\frac{A(A\to B)}{A(B\to A)}
=e^{-\beta (E_B-E_A)}\,.
\label{detbalbol2}
\end{equation}
Noticing that $E_A-E_B=2J(n-m)$, it follows that:
\begin{equation}
\frac{A(A\to B)}{A(B\to A)}=\left[(1-P_{\rm add})e^{2\beta J}\right]^{n-m}\,.
\end{equation}
Therefore, choosing $P_{\rm add}=1-e^{2\beta J}$, the acceptance
probabilities simplifies: $A(A\to B)=A(B\to A)=1$. For this reason the spins
can be automatically flipped when the cluster is formed. In the vicinity
of the critical point, the Wolff algorithm significantly reduces the
autocorrelation time and the critical dynamical exponent compared to
a local algorithm (such as Metropolis or Glauber). We notice
that the time $\tilde\tau_W$ measured in units of Wolff iterations
involves a subset of spins corresponding to the averaged size $\langle
p\rangle$ of a cluster. On the other hand a time $\tau_M$ measured in
units of Metropolis iterations involves all spins of the network
i.e. $N=L^D$ spins. To compare the efficiency of both algorithms fairly, it is
therefore necessary to define a rescaled
Wolff autocorrelation time $\tau_W=\tilde\tau_W\langle
p\rangle/L^D$. Moreover, it is possible to show that $\chi=\beta\langle
p\rangle$~\cite{Newman99}. Noticing that $\tilde\tau_W\sim L^{\tilde
z^W_c}$, it follows (remembering $\xi\sim L$) $\tau_W\sim\xi^{z^W_c}\sim
L^{\tilde z^W_c+\gamma/\nu-D}$ leading to the definition of the
dynamical critical exponent  $z^W_c=\tilde z^W_c+\gamma/\nu-D$. In
2$D$ for example, remarkably, the dynamical exponent is $z^W_c\approx0$
for Wolff (see e.g.~\cite{Gunduc05,Du06} and references therein)
whereas $z^M_c=2.1665(12)$~\cite{Nightingale96} for Metropolis.

\section{Continuous-time or rejection free algorithm}
As we have seen in the previous subsection, with a local algorithm
(like e.g. Metropolis) a spin flip of the Ising model at criticality
has a high probability to be rejected, and this holds even more in the
ferromagnetic phase. A
significant amount of the computational time will therefore be spent
without making the system evolve. An alternative way has been proposed
by Gillespie~\cite{Gillespie76} in the context of chemical reactions and
afterwards applied by Bortz, Kalos and Lebowitz in the context of spin
systems~\cite{bkl}.

Briefly, this algorithm lists all possible Monte Carlo moves that can
be performed in the system in its current configuration. One of these moves
is chosen randomly according to its probability, and the system is forced
to move into this state. The time step of evolution during such a move can
be estimated rigourously. This time will change from each configuration
and cannot be set to unity as in the Metropolis algorithm: it takes
a continuous value. this is why this algorithm is sometimes called
{\it continuous time algorithm}. On the one hand, this algorithm has to
maintain a list of all possible moves, which requires a relatively heavy
administrative task, on the other hand, the new configuration is always
accepted and this saves a lot of time when the probability of rejection
would otherwise be high. It is also sometimes called the {\it rejection
free algorithm}. The efficiency of this algorithm will be maximized for
$T\le T_c$. In detail, one iteration of the continuous time algorithm looks like:
\begin{enumerate}
 \item List all possible moves from the current configuration. Each of
  these $n$ moves has an associated probability $P_n$. 
 \item Calculate the integrated probability that a move occurs $Q=\sum_{i=0}^n P_n$.
 \item Generate a random number $\rm{\bf{rn_1}}$ uniformly distributed in $[0,Q]$.
 This selects the chosen move with probability $P_n/Q$.
 \item estimate the time elapsed during the move:
 $\Delta t=Q^{-1}\ln\,(1-\rm{\bf{rn_2}})$ where $\rm{\bf{rn_2}}$ is a random number uniformly
 distributed in $[0,1]$ \footnote{The probability of a spin flip is
 exponential versus $Q$: $P(\Delta t)=\exp(-Q\Delta t)$).}
\end{enumerate}

Implementation of this algorithm becomes easier if the probabilities $P_n$
can only take a small number of values. In that case, lists can be made of
all moves with the same probability $P_n$. The selection process is
then first to select one of the lists, with the appropriate probability,
after which randomly one move is selected from that list.  This is the
case e.g. in Ising simulations on a square (2$D$) or cubic (3$D$) lattice, when the probability
$P_n$ is limited to the values $1$, $e^{-4\beta J}$, $e^{-8\beta J}$, or $e^{-12\beta J}$
(the latter occurring only in 3$D$).

\section{The worm algorithm} 

We present here another example of a local algorithm, the so-called
{\it worm algorithm} introduced by Prokof'ev, Svistunov and
Tupitsyn~\cite{Pro&Svi&Tup,Pro&Svi}. The difference with the algorithms
presented above is an alternative representation of the system, in
terms of graphs instead of spins. The Markov chain is therefore
performed along graph configurations rather than spin configurations,
but always with Metropolis acceptance rates. The principle is based on
the high temperature expansion of the partition function. Suppose that we want
to sample the magnetic susceptibility of the Ising model. We can access 
it via the correlation function using the (discrete)
fluctuation-dissipation theorem:
\begin{equation}
\chi=\frac{\beta}{N} \sum_{i,j} G(i-j)\,,
\label{FDT}
\end{equation}
where $G(i-j)=\langle S_{i}\cdot S_{j}\rangle-\langle S_{i}\rangle^2$
is the connected correlation function between sites $i$ and $j$. In
the high temperature phase, the average value of the spin cancels and
$G(i-j)=\langle S_{i}S_{j}\rangle$. The first step is to write the
correlation function $G(i-j)$ of the Ising model in the following form:
\begin{eqnarray}
G(i-j)&=&
\frac{1}{Z}\sum_{\{S\}}S_{i}\cdot S_{j}\,e^{-\beta J \sum_{\langle k,l\rangle}S_{k}\cdot S_{l}}\,,\label{Dev1}\\
 &=&\frac{1}{Z}\cosh(\beta J)^{DN}\sum_{\{S\}}S_{i}\cdot S_{j}\prod_{\left< k,l\right>}\left(1+S_{k}\cdot S_{l}\tanh(\beta J)\right)\,.\label{Dev2}
\end{eqnarray}
\begin{figure}[!ht]
\begin{center}
\includegraphics[width=0.8\columnwidth]{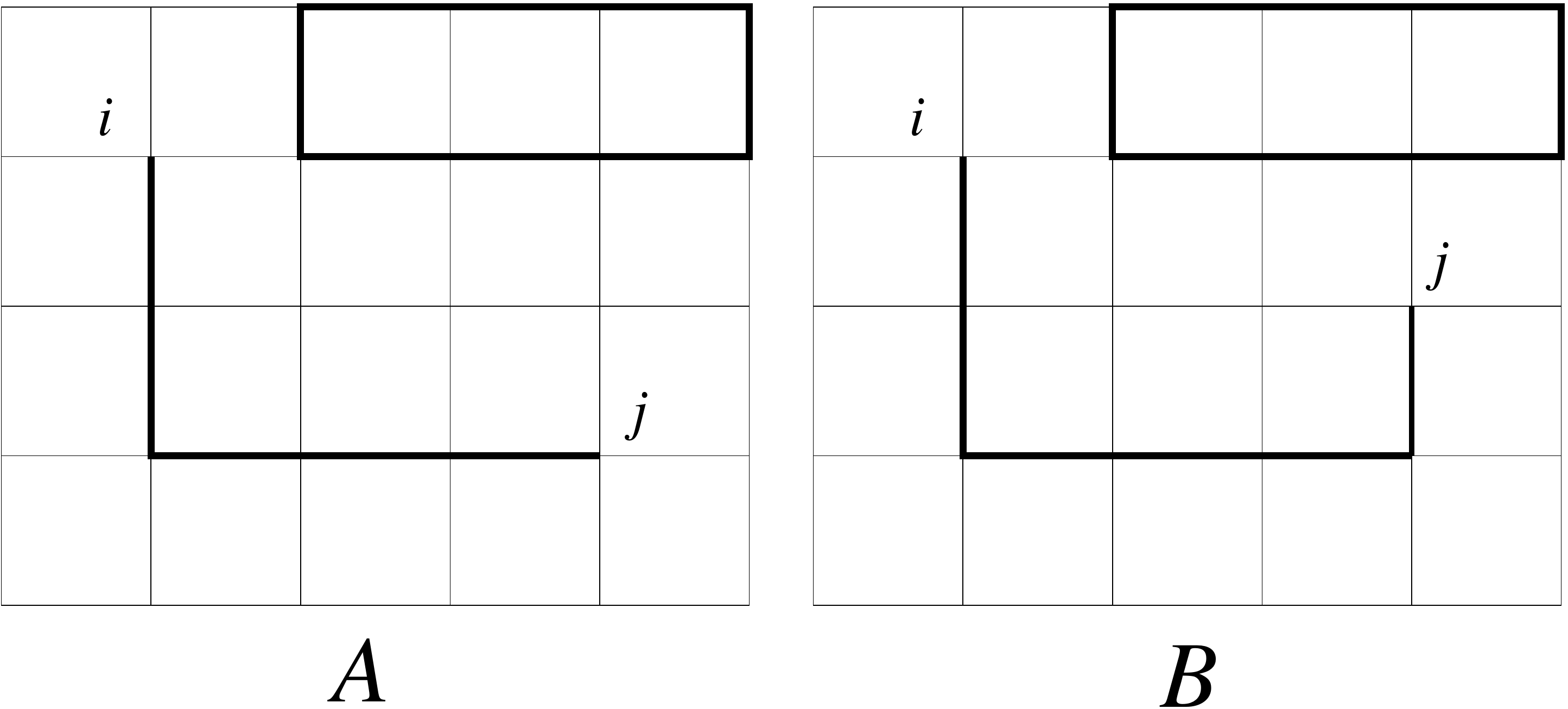}
\caption{\label{Worm-Bis} Illustration of a move with the worm
algorithm. Thick lines stand for one example of graph contributing
to the correlation function: one path joining the sites $i$ et $j$
(the sources) and possibly closed loops. These two graphs differ in one
iteration of the worm algorithm. According to (\ref{Dev2}), the graph
on the left and on the right have respective equilibrium probabilities
$P_A\propto\tanh^5\beta J$ and $P_B\propto\tanh^6\beta J$ (we neglect
loops that are not relevant for this purpose). From the detailed balance
condition (\ref{detbal}), the Metropolis acceptance rates are $A(A\to
B)={\rm Min}(1,\tanh\beta J)$ and $A(B\to A)={\rm Min}(1,1/\tanh\beta
J)$. In both case $T(A\to B)=T(B\to A)=1/2D$ where $D$ is the dimension
of the (hypercubic) lattice.}
\end{center}
\end{figure}
The configurations that contribute to the sum in (\ref{Dev2}) contain
an even number of spins in the product at any given site. Other products
involving an odd number of spins in the product contribute zero. Each
term can be associated with a path determined by the sites involved in
it. A contribution to the sum is made of a (open) path joining sites $i$
and $j$ and closed loops. The sum over the configurations can be replaced
by a sum over such graphs. Figure~\ref{Worm-Bis} sketches such a
contribution for a given couple of source sites $i$ and $j$.
The importance sampling is no longer made over spin configurations but
over graphs that are generated as follows. One of the two sources, say $i$,
 is mobile. At every steps,
it moves randomly to a neighboring site $n$. Any nearest-neighbor site
can be chosen with the trial probability $T(A\to B)=1/2D$, where $D$
is the dimension of the (hypercubic) lattice. If no link is present between
the two sites, then a link is created with the acceptance probability:
\begin{equation}
A(A\to B)={\rm Min}(1,\tanh\beta J)\,.
\label{taux1}
\end{equation}
If a link is already present, it is erased with the acceptance probability:
\begin{equation}
A(A\to B)={\rm Min}(1,1/\tanh\beta J)\,.
\label{taux2}
\end{equation}
Since $0\le\tanh x<1$ for all values $x>0$, the probability (\ref{taux2}) is
equal to unity and the link is always erased. These probabilities are
obtained considering the Metropolis acceptance rate Eq.(\ref{metro}) and
the expression of the correlation function (\ref{Dev2}). The procedure
is illustrated in Figure~\ref{Worm-Bis}.

The open paths in the two graphs are resp. made of 5 and 6
lattice spacings (we neglect the loop that does not contribute in this
example). According to (\ref{Dev2}), the graphs on the left and on the
right have equilibrium probabilities $P_A\propto\tanh^5(\beta
J)$ and $P_B\propto\tanh^6(\beta J)$, respectively. The transition
probability (\ref{metro}) is thus written $W(A\to B)=1/2D\times{\rm
Min}(1,\tanh\beta J)$ and $W(B\to A)=1/2D\times{\rm Min}(1,1/\tanh\beta
J)$, in agreement with (\ref{taux1}) and (\ref{taux2}).

\begin{figure}[!ht]
\begin{center}
\centerline{\hbox{
\includegraphics[width=0.6\columnwidth]{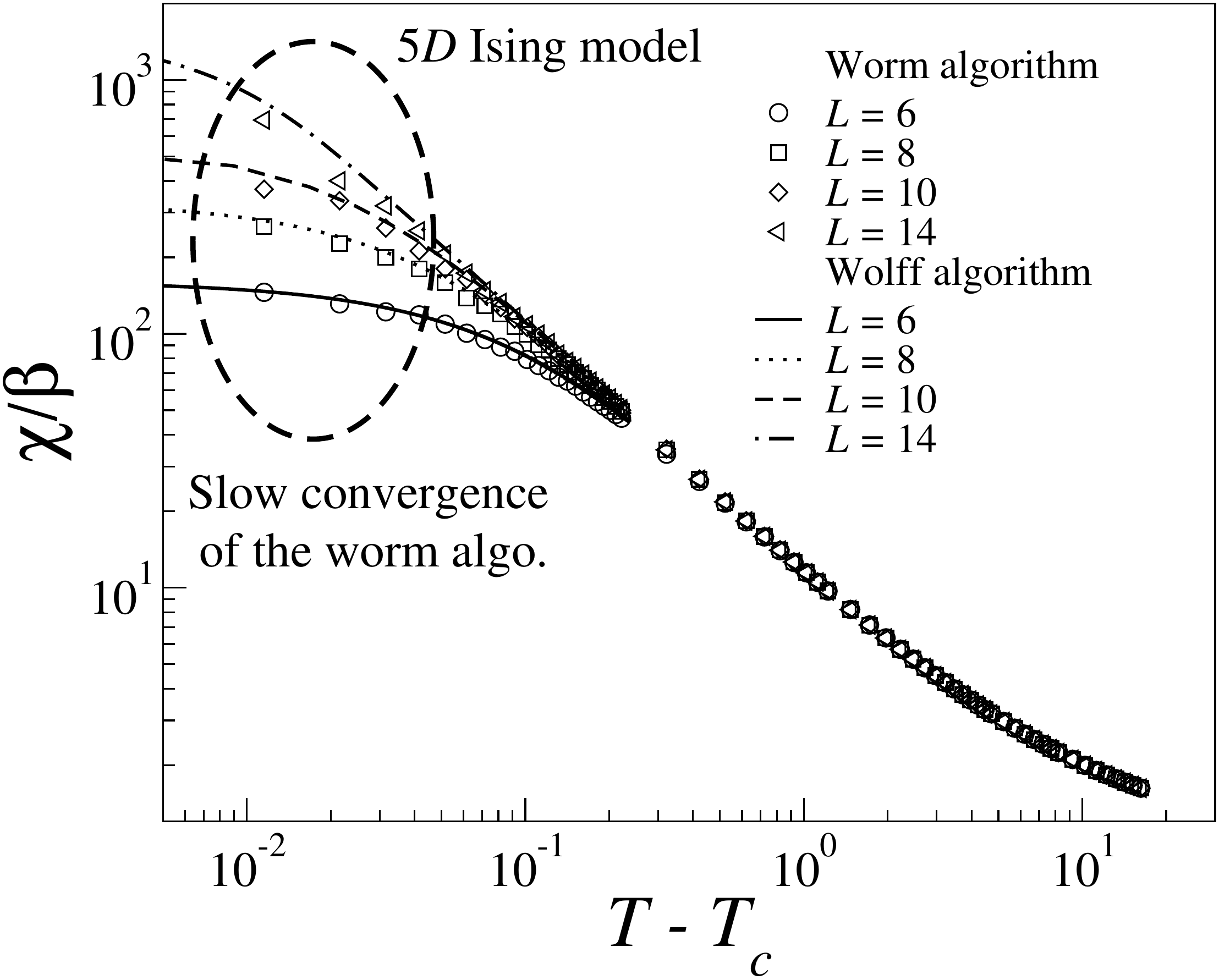}
}}
\caption{\label{test} Comparison of the magnetic susceptibility $\chi$
obtained with the worm algorithm and the Wolff algorithm for the Ising
model in 5$D$ with different lattice sizes. The results are in excellent
agreement for both algorithms in the high temperature phase. As we
move closer to $T_c$ i.e. in the critical regime (dashed line ellipse),
the convergence of the worm algorithm is slower as the lattice size increases 
(keeping all other parameters fixed). The number of graphs increases exponentially.}
\end{center}
\end{figure}

If the two sources meet, they can move together on another random site
with a freely chosen transition probability. When the two sources move
together, they leave a closed loop behind that justifies the simultaneous
presence of open path and closed loop in Figure~\ref{Worm-Bis}. These
loops may disappear if the head of the worm meets them. Compared to
the Swendsen-Wang algorithm, the worm algorithm has a dynamical
exponent slightly higher in 2$D$ but significantly lower in
3$D$~\cite{Deng07}. The efficiency of this algorithm can be improved
with the use of a continuous time implementation~\cite{berche08}. The
formalism of the worm algorithm is suitable for high temperature. In the
critical region, the number of graphs that contribute to the correlation
function increases exponentially. In order to check the convergence of
the algorithm, we can compare it with the Wolff algorithm for the 5$D$
Ising model with different lattice sizes in Figure~\ref{test}. The
two algorithms give results in good agreement, except in the critical
region where the converge of the worm algorithm is slower as the lattice size $L$ increases.

\section{Dynamical aspects: thermalization \& correlation time}

\begin{figure}[!ht]
\begin{center}
\centerline{\vbox{
\includegraphics[width=0.9\columnwidth]{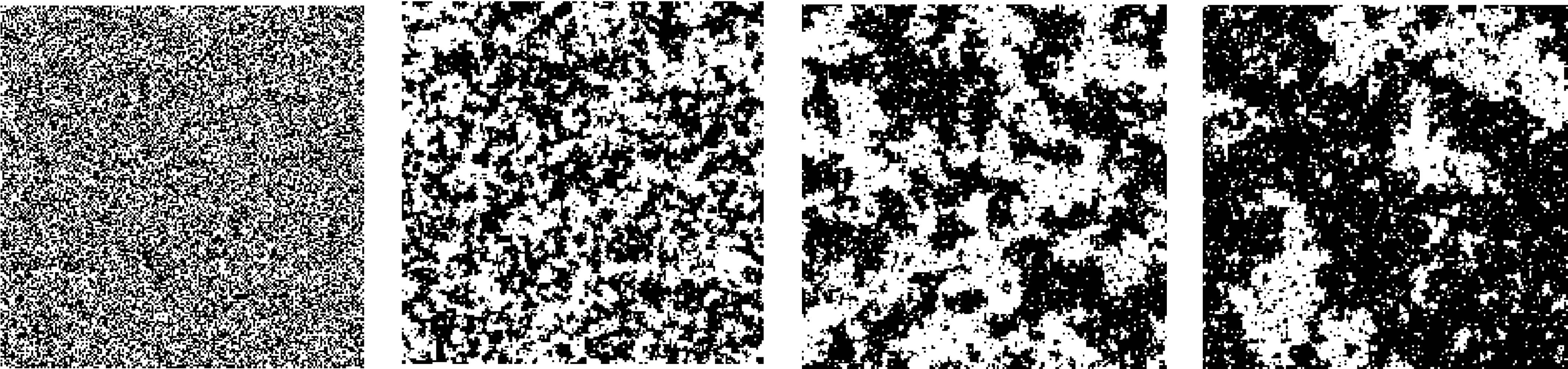}
}}
\caption{\label{quench} Snapshots of the evolution of the 2$D$ Ising of size
$200\times200$ after a quench from a disordered state until equilibration
at the critical temperature $T_c$ with the Glauber dynamics. We see the
nucleation and the growth of correlated domains. From left to right,
$t=$0, 10, 100, 1000 (expressed in Monte Carlo unit time after the
quench). The thermalization is completed when the correlation length
reaches its static value $\xi\sim L$. Using Eq.(\ref{xi-crit}), the
thermalization time at criticality behaves like $\tau_{\rm th}\sim
L^{z_c}$ and therefore diverges with system size.}
\end{center}
\end{figure}

In order to perform a sampling of thermodynamical quantities at a given
temperature, one has to first thermalize the system. Usually, it is
possible to set up the system either at infinite temperature (all spins
random) or in the ground state (all spins up or down). Let us start from
an initial random configuration. If the thermalization takes place above
the critical temperature, then the relaxation is exponential. As we come
closer to the critical point, the equilibrium correlation length becomes
larger and the relaxation becomes much slower and eventually algebraic
right at $T_c$. An example of such process for the 2$D$ Ising model is
given in Fig.~\ref{quench}. Initially ($t=0$), the system is prepared
at infinite temperature, all the spins are random. Then the Glauber dynamics
 is applied at the critical temperature. We see the nucleation
and the evolution of correlated spin domains in time ($t=$0, 10, 100,
1000 from left to right). It is possible to show that in such a quench,
the correlation length grows with time like~\cite{Bra94}:
\begin{equation}
\xi(t)\sim t^{1/z_c}\,,
\label{xi-crit}
\end{equation}
where $z_c$ is the critical dynamical exponent
($z_c=2.1665(12)$~\cite{Nightingale96} for the 2$D$ Ising model). The time needed
to complete thermalization at criticality is therefore $\tau_{\rm
th}\sim L^{z_c}$. In case of a subcritical quench, the system has to
choose between two ferromagnetic states of opposite magnetization. Again,
the relaxation is slow because there is nucleation and growth of domain of
opposite magnetization. We define the typical size of a domain at a time $t$ by $L_d(t)$. 
The thermalization process involves the growth
(coarsening) of these domains, until eventually one domain spans the whole
system. Only then, equilibrium is reached (in the low temperature
phase, the expectation of the absolute value of the magnetization is nonzero). 
The motion of the domain walls is mostly diffusive $L_d(t)\sim t^{1/z}$ with
a dynamical exponent $z=2$~\cite{Bra94}. The walls have to cover a distance $\sim L$,
so that the thermalization time scales as $\tau_{\rm th} \sim L^2$. This time diverges
again with system size. Starting from an ordered state does not help for the critical
quench (but it does help to start in the ground state to thermalize the system at $T<T_c$).
 
Once the system is thermalized, one has to be aware of another dynamical
effect: the correlation time. This is the time needed to perform sampling
between statistically uncorrelated configurations. In the high temperature phase, the 
correlation time is equal to the thermalization time, up to some factor
close to unity. This is not surprising, as proper thermalization requires
the configuration to become uncorrelated from the initial state.
Practically, in all Monte Carlo simulations, one has to estimate $\tau$
at the temperature of sampling to treat properly the error bars.

In the low temperature phase, after thermalization, the magnetization is either positive or
negative, and stays like that over prolonged periods of time. So-called
magnetization reversals do occur now and then, but the characteristic
time between those increases exponentially with system size. Because of
the strict symmetry between the parts of phase space with positive and
negative magnetization, in practice one is not so much interested in the
time of magnetization reversals, but rather in the correlation time $\tau$
within the up- or down-phase; and this time is some temperature-dependent
constant, irrespective of the system size provided it is significantly
larger than the correlation length.

Let us consider now the two-time spin-spin correlation functions in the
framework of dynamical scaling~\cite{hohenberg_theory_1977}. We will 
use for this purpose a continuous space, so that the spin $S_i$ on the site $i$ is 
now denoted by $S_{\vec r}$ where $\vec r$ is the position vector. 
Upon a dilatation with a scale factor $b$, the equilibrium correlation
$C(\vec r,t,|T-T_c|)=\langle S_0(0)\cdot S_{\vec r}(t)\rangle$ 
is assumed to satisfy the homogeneity relation:
      \begin{equation}
        C(\vec r,t,|T-T_c|)=b^{-2x_\sigma}C\big(r/b,t/b^{z_c},|T-T_c|b^{1/\nu}\big)\,,
        \label{eq1c}
        \end{equation}
where $x_\sigma$ is the scaling dimension of magnetization density with
$2x_\sigma=\eta$ for two-dimensional systems and $z_c$ is again the critical dynamical
exponent. The motivation for the last two arguments of the scaling function
in equation (\ref{eq1c}) comes from the behavior of the correlation length
either with time, $\xi\sim t^{1/z_c}$, or with temperature,
$\xi\sim |T-T_c|^{-1/\nu}$. Setting $b=t^{1/z_c}$ in equation (\ref{eq1c}),
we obtain:
       \begin{equation}
         C(\vec r,t)=t^{-\eta/z_c}{\cal C}\big(r/t^{1/z_c},|T-T_c|t^{1/(\nu
           z_c)}\big)\,.\label{ScalingHyp}
       \end{equation}
The algebraic prefactor corresponds to the critical behavior while
the scaling function includes all corrections to it. The characteristic time:
        \begin{equation}
          \tau\sim \xi^{z_c}\sim|T-T_c|^{-\nu z_c}\,,
          \label{ScalTau}
          \end{equation}
appears as the relaxation time of the system. Here we are
interested only in autocorrelation functions for which $r=0$. Moreover, we
expect an exponential decay of the scaling function ${\cal C}(t/\tau)$
in the paramagnetic phase. Therefore, the autocorrelation function can generally
be written at equilibrium as:
       \begin{equation}
         C(t,T)\sim {\frac{e^{-t/\tau}}{t^{\eta/z_c}}}\,.
         \label{eq2}
       \end{equation}

\begin{figure}[!ht]
\begin{center}
\centerline{\hbox{
\includegraphics[angle=0,width=0.5\columnwidth]{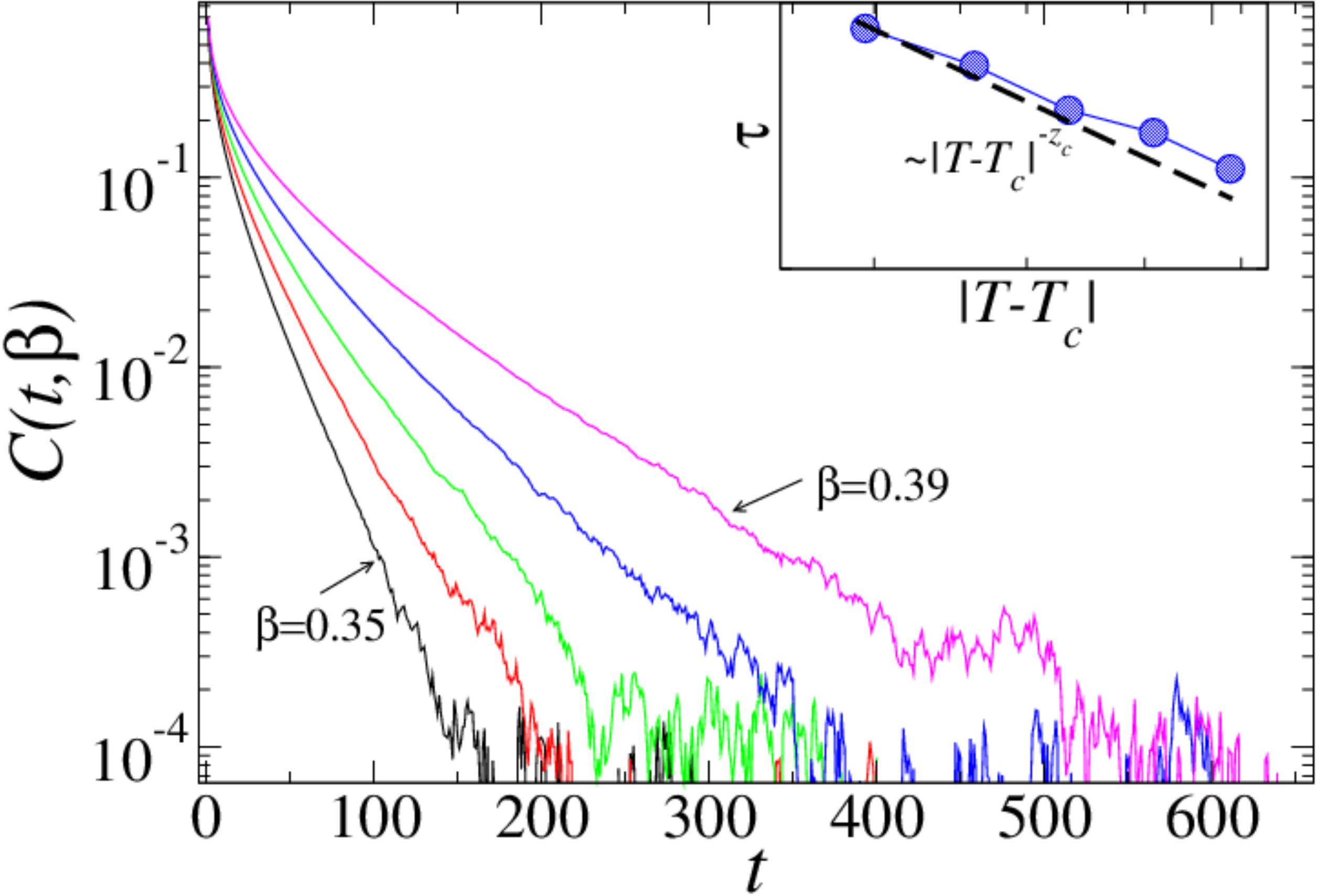}
\includegraphics[angle=0,width=0.5\columnwidth]{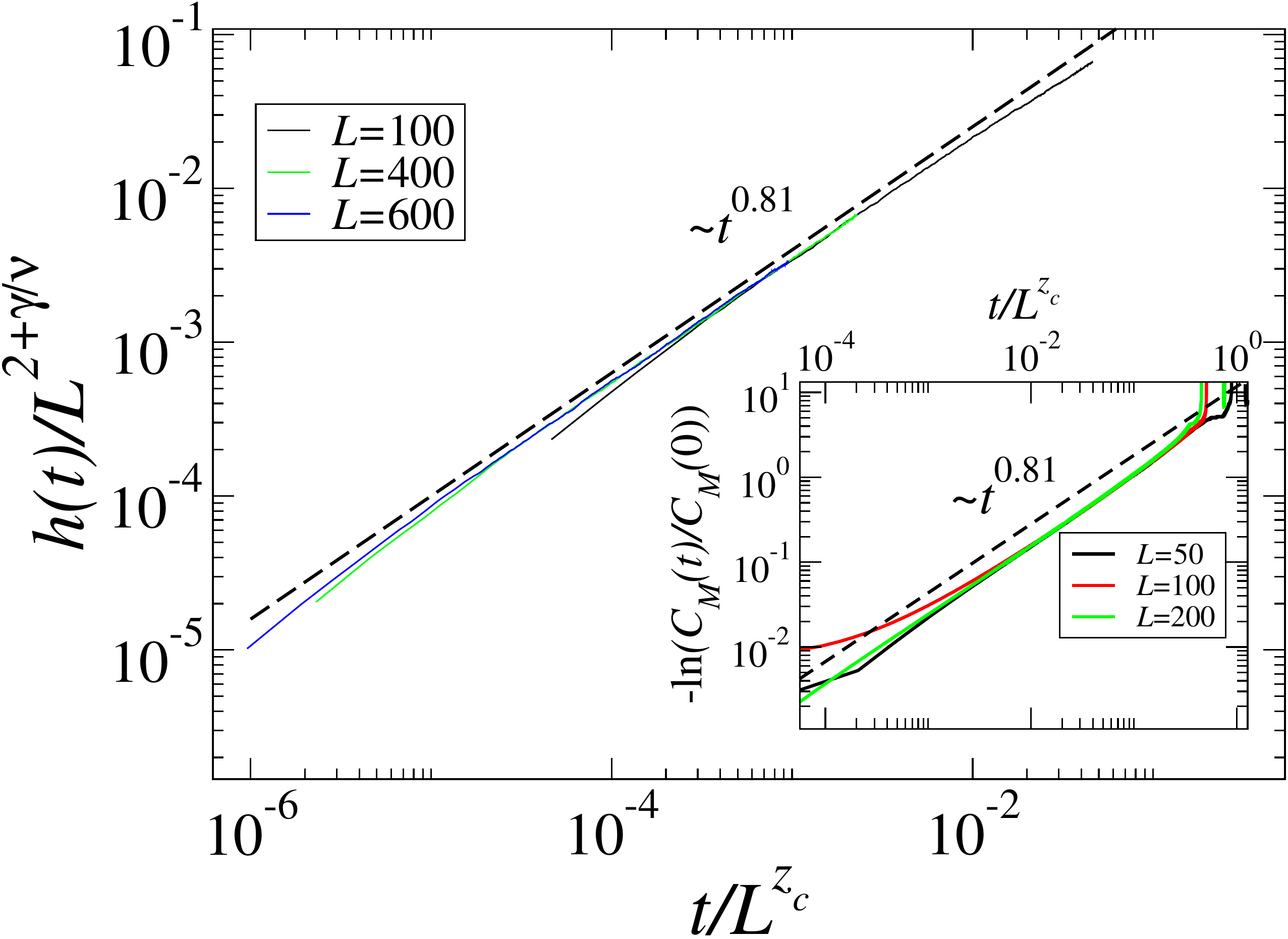}
}}
\caption{\label{fig:mcor}(left) Spin-spin autocorrelation function $C(t)$ of the 2$D$ Ising model
($N=50\times50$) at different $\beta=J/(k_BT)=$0.35, 0.36, 0.37, 0.38 and 0.39 ($\beta_c\approx0.44$). 
The correlation time $\tau$ is an increasing function of $T-T_c$. Assuming Eq.(\ref{eq2}),
$\tau$ that is plotted in the inset vs. $|T-T_c|$.
(right) Scaling function $h(t)/L^{2+\gamma/\nu}$ of the mean-square deviation of the magnetization
versus $t/L^{z_c}$ for the 2$D$ Ising model at $T_c$ for different lattice sizes. 
At intermediate times, it displays an {\it anomalous diffusion} behavior compatible with $h(t)\sim t^{\gamma/(\nu z_c)}=t^{0.81}$. 
Insert: the autocorrelation $C_M(t)=\left<|M(0)|\cdot|M(t)|\right>$ is compatible with 
a stretched exponential $C_M(t)\sim\exp\left(-(t/\tau)^{\gamma/(\nu z_c)}\right)$, explained 
by the behavior of $h(t)$.
}
\end{center}
\end{figure}

The spin-spin autocorrelation function $C(t,T)$ versus time is plotted in Fig.\ref{fig:mcor} (left) for the 
2$D$ Ising model of size $N=50\times50$. The different curves correspond to different inverse 
temperatures $\beta=0.35$ to 0.39 (the critical inverse temperature is $\beta_c=1/2\ln(1+\sqrt 2)\approx0.441$). We observe 
an increase of the autocorrelation time as the temperature comes closer to $T_c$. The autocorrelation time
can be obtained from a fit of the curve in the main graph, assuming Eq.(\ref{eq2}). The result is plotted versus $|T-T_c|$
in the inset. The numerics tend to the behavior of Eq.(\ref{ScalTau}) as $T$ goes to $T_c$.
       
Some other aspects of critical dynamics are interesting to study, for instance, the time evolution of the equilibrium mean-square
displacement of the magnetization. It is defined as:
\begin{equation}
h(t) = \left< \left( M(t) - M(0)\right)^2\right>.
\label{eq:mcor}
\end{equation}

At small time differences ($t < 1$), the dynamics consists of sparsely
distributed proposed spin flips, each of which has a nonzero acceptance probability.
Since these spin flips are uncorrelated and their number scales as $L^D t$, in the short-time
regime ($t\approx1$), $h(t)$ behaves diffusively:
\begin{equation}
h(t) \sim L^D t\,.
\label{eq:early}
\end{equation}

The magnetization at long time $t> \tau\sim L^{z_c}$ is no longer correlated 
i.e. $\left< M(t) \cdot M(0) \right> \approx 0$.  Moreover, the expectation
value of the squared magnetization is directly related to the magnetic susceptibility like
$\chi \equiv \beta/N \left<M^2\right>$ where $N=L^D$ (for $T\ge T_c$, $\left<M\right>\approx0$).
It diverges at the critical temperature with system size as $\sim L^{\gamma/\nu}$, implying:
\begin{equation}
h(t) = \left< M(t)^2 + M(0)^2 -2 M(t)\cdot M(0)\right> \approx 2 \left<M^2\right> \sim L^{D+\gamma/\nu}
\hspace{1cm} (t>\tau).
\end{equation}
Therefore $h(t)$ has to grow from $h(t\approx1)\sim L^D$ to $h(t\sim L^{z_c})\sim L^{D+\gamma/\nu}$. 
Assuming a power law behavior, it follows that $h(t)\sim t^{\gamma/(\nu z_c)}$.
Therefore, in this regime, we can assume the following form for $h(t)$:
\begin{equation}
h(t)\sim L^{D+\gamma/\nu}\mathcal F(t/L^{z_c})\,,\label{F}
\end{equation}
where $\mathcal F$ is a scaling function with the limit $\mathcal F(x)={\rm constant}$ when $x\gg1$ and
$\mathcal F(x)\sim x^{\gamma/(\nu z_c)}$ at intermediate times. 
We measured the function $h(t)$ as defined in Eq.(\ref{eq:mcor}) in simulations of the two-dimensional Ising model
at the critical temperature for various systems sizes. The scaling function $\mathcal F$ is plotted 
in Fig.~\ref{fig:mcor} (right), using the exponents $\gamma=1.75$, $\nu=1$ and $z_c\approx2.17$.
With increasing system size, the data become increasingly consistent with a simple
power law behavior (for intermediate times between the early-time behavior (Eq.~(\ref{eq:early}) and the time of saturation $\sim L^{z_c}$).
This power law behavior corresponds to an instance of {\it anomalous diffusion} i.e. a mean-square deviation growing
as a power law with an exponent $\ne1$ compatible with:
\begin{equation}
h(t) \sim t^{\gamma/(\nu z_c)} \sim t^{0.81}.\label{AD}
\end{equation}

Assuming Eq.(\ref{AD}), the magnetization autocorrelation function for intermediate times (i.e. between times
of order unity and the correlation time, thus spanning many decades) is compatible with the first terms of the 
Taylor expansion of a stretched exponential:
\begin{equation}
C_M(t)=\left< |M(t)|\cdot |M (0)|\right> \sim \exp \left( -(t/\tau)^{\gamma/(\nu z_c)}\right).
\end{equation}
This conjecture compares well with the numercis for the correlation function in Figure~\ref{fig:mcor} (right, in the inset).
This shows that the dynamical critical exponent $z_c$ appears at relatively small
times $1 \ll t \ll L^{z_c}$.

\section{Conclusion}

In these lecture notes, we provide an introduction to Monte Carlo simulations that are a way to produce a set 
of representative configurations of a statistical system. We start with the basic principles: ergodicity and detailed balance. 
In the next parts, we present several Monte Carlo algorithms. To illustrate their functioning, we use the example of the Ising model. 
This model is defined by scalar spins on a lattice that interact via nearest-neighbor interactions. This is the paradigmatic 
system for second order phase transitions: a critical temperature shares a disorderered phase at high temperature 
and an ordered phase at low temperature. These different regimes induce different strategies for the Monte Carlo simulations. 
In the disordered phase, local algorithms such as Metropolis or Glauber are efficient. In the critical region, 
the appearance of long range correlations have set a computational challenge. It has been solved
by the use of cluster algorithms such as the Wolff algorithm that flips a whole cluster of correlated spins.
Below the critical temperature, when the probability of spin flip is low, it is a gain of computational time
to program the continuous time algorithm. It forces the system into a new configurations, with a jump in time
according to the probability of transition. Finally, we describe an interesting algorithm based on an alternative
representation of the model in terms of graphs instead of spins. We end with important considerations on the dynamics:
thermalization and correlation time.
       
\section*{Acknowledgements}

We thank Christophe Chatelain for his careful reading of the manuscript and
 the various collaborations that have largely inspired these notes.
 We also thank Raoul Schram for stimulating discussions and the reading of the manuscript.
 J-CW is supported by the Laboratory of Excellence Initiative (Labex) NUMEV, OD by the Scientific Council
 of the University of Montpellier 2.
 This work is part of the D-ITP consortium, a program of the Netherlands Organisation for Scientific Research (NWO)
that is funded by the Dutch Ministry of Education, Culture and Science (OCW).





\end{document}